# Physics of the mind:
## Concepts, emotions, language, cognition, consciousness, beauty, music, and symbolic culture

Leonid I. Perlovsky

Harvard University and Air Force Research Lab.

**Abstract**

Mathematical approaches to modeling the mind since the 1950s are reviewed. Difficulties faced by these approaches are related to the fundamental incompleteness of logic discovered by K. Gödel. A recent mathematical advancement, dynamic logic (DL) overcame these past difficulties. DL is described conceptually and related to neuroscience, psychology, cognitive science, and philosophy. DL models higher cognitive functions: concepts, emotions, instincts, understanding, imagination, intuition, consciousness. DL is related to the knowledge instinct that drives our understanding of the world and serves as a foundation for higher cognitive functions. Aesthetic emotions and perception of beauty are related to 'everyday' functioning of the mind. The article reviews mechanisms of human symbolic ability, language and cognition, joint evolution of the mind, consciousness, and cultures. It touches on a manifold of aesthetic emotions in music, their cognitive function, origin, and evolution. The article concentrates on elucidating the first principles and reviews aspects of the theory proven in laboratory research.

### Physics and the mind

Physics concentrates on the fist principles, a few fundamental laws of nature explaining significant part of knowledge in a wide field. Recent discoveries in neuroscience make it possible to identify the first principles of the mind-brain. These fundamental laws are the subject of this article.

### Logic vs. mind

For a long time people believed that intelligence is equivalent to logical conceptual understanding and reasoning. Although it is obvious that the mind is not logical, over the course of the two millennia since Aristotle, many people have identified the power of intelligence with logic. Founders of artificial intelligence in the 1950s and 60s believed that by relying on rules of logic they would soon develop computers with intelligence far exceeding the human mind.

This story begins with Aristotle, the inventor of logic[1]. Aristotle, however, did not think that the mind works logically; he invented logic as a supreme way of argument, not as a theory of the mind. This is clear from his many writings, for example, in "Rhetoric for Alexander" Aristotle lists dozens of topics on which Alexander had to speak publicly[2]. For each topic, Aristotle identified two opposite positions (e.g. make peace or declare war; use torture or offer gold for extracting the truth, etc.). For each of the opposite positions, Aristotle gave logical arguments to argue either way. Clearly, for Aristotle, logic is a tool to express previously made decisions, not the mechanism of the mind. Logic can only provide deductions from the first principles, but cannot indicate what the first principles should be. To explain the mind, Aristotle developed a theory of Forms, which will be discussed later. But during the following centuries the subtleties of Aristotelian thoughts were not always understood. With the advent of science, the idea that intelligence is equivalent to logic was gaining grounds. In the 19th century mathematicians turned their attention to logic. George Boole noted that foundations of the Aristotelian theory of logic were unsatisfactory. These foundations included the law of excluded middle (or excluded third) stating that every statement is either true or false and any alternative is excluded[3]. But Aristotle also emphasized that logical statements should not be formulated too precisely (say, a measure of wheat should not be defined with an accuracy of a single grain), that language implies the adequate accuracy, and everyone has his mind to decide what is the reasonable accuracy.

Boole thought that the contradiction between the exactness of the law of excluded middle and the vagueness of language should be corrected. A new branch of mathematics, *formal logic* was born. Prominent mathematicians contributed to the development of formal logic, including George Boole, Gottlob Frege, Georg Cantor, Bertrand Russell, David Hilbert, and Kurt Gödel. Logicians 'threw away' uncertainty of language and founded formal mathematical logic based on the law of excluded middle. Hilbert developed an approach named formalism, which attempted to define scientific objects formally in terms of axioms or rules. Hilbert was sure that his logical theory also described mechanisms of the mind: "The fundamental idea of my proof theory is none other than to describe the activity of our understanding, to make a protocol of the rules according to which our thinking actually proceeds."[4]

Almost as soon as Hilbert formulated his formalization program, the first hole appeared. In 1902 Russell exposed an inconsistency of formal logic by introducing a set R as follows: *R is a set of all sets which are not members of themselves.* Is R a member of R? If it is not, then it should belong to R according to the definition, but if R is a member of R, this contradicts the definition. Thus, either way we get a contradiction. This became known as the Russell's paradox. Its colloquial formulation asks the following question. "A barber shaves everybody who does not shave himself. Does the barber shave himself?" Either answer to this question (yes or no) leads to a contradiction. This barber, like Russell's set, can be logically defined, but cannot exist. For the next 30 years mathematicians where trying to develop a self-consistent mathematical logic, free from the paradoxes of this type. But, in 1931, Gödel has proved that it is not possible[5], formal logic was inconsistent, self-contradictory.

Belief in logic has deep psychological roots related to functioning of human mind. A major part of perception and cognition is not accessible to consciousness directly. We are not conscious of neural firings; we are conscious about the 'final states' of these processes, which are perceived by our minds as 'concepts' approximately obeying formal logic. For this reason lay people and prominent mathematicians believe in logic. Even after the Gödelian proof, founders of artificial intelligence in the 1950s and 1960s still insisted that logic is sufficient to explain working of the mind.

### Difficulties of modeling the mind since the 1950s: Complexity and logic

Simple object perception involves "bottom-up" signals from sensory organs and "top-down" signals from mental representations (memories) of objects. During visual perception these signals interact in the visual cortex, the mind associates signals from objects with mental representations of object. This produces object recognition; it activates brain signals leading to mental and behavioral responses, which constitutes understanding.

Developing mathematical descriptions of the very first *recognition* step in this seemingly simple association-recognition-understanding process met irresolvable difficulties. These difficulties were summarized under the notion of combinatorial complexity (CC)[6]. CC refers to multiple combinations; say recognition of a scene requires concurrent recognition of multiple objects that could be encountered in various combinations. CC is prohibitive because the number of combinations is very large: for example, consider 100 objects (not too large a number, when you look in any direction you often see more than 100 objects). The number of combinations of 100 objects is $100^{100}$, exceeding the number of all elementary particle events in life of the Universe. The mind would never be able to compute that many combinations.

It turned out that various manifestations of CC in artificial intelligence, pattern recognition, neural networks, fuzzy logic, etc. are all related to formal logic and Gödel theory. Even mathematical approaches specifically designed to overcome logic, like neural networks and fuzzy logic, still rely on logic during training or learning procedures: e.g., "this is a chair" – is a logical statement.

### Dynamic logic (DL)

DL was invented to overcome difficulties of classical formal logic and CC[7],[8]. According to DL, mental representations in memory are vague and approximately correspond to multiple objects and situations. Therefore there is no need to consider multiple combinations. In processes of perception and cognition, mental representations are modified to better fit sensor data. As fits improve, vagueness is reduced. Representations become crisper; they compete for evidence in data (bottom-up signals). Representations that best fit data, win the competition, become crisp and available to consciousness. To summarize, DL is a process from-vague-to-crisp, and from unconscious to conscious (or from less conscious to more conscious).

Relying on knowledge of neural mechanisms of perception discovered in a recent decade, everyone can experimentally verify in 3 seconds that the DL vague-to-crisp process operates in one's brain-mind. Close your eyes and imagine an object in front of you. This imagination is vague, not as crisp as perception with opened eyes. Also, it is not as conscious, and with opened eyes, it is difficult to consciously remember the imagination. Similar experiment was performed with much more details using brain imaging at Harvard University in the Laboratory of Moshe Bar[9]. Perception of objects is not momentary as it may seem. The process takes approximately $1/6^{th}$ sec. Conscious perceptions are preceded by activations of cortex areas storing memories-representations of objects. The initial projections of these representations to visual cortex are *vague*. These vague representations and the entire $1/6^{th}$ sec perception process are not accessible to consciousness.

Example of DL operations is illustrated in Fig. 1. The true patterns without noise are shown in (A), and the actual data with noise are in (B). Previously, finding patterns under noise, like in (B), was an unsolvable problem because of CC of fitting models to the data. DL solves this problem as illustrated in (C) through (H). Beginning with a vague model (representation) in (C) it converges to crisp ones in (H) similar to the true patterns in (A).

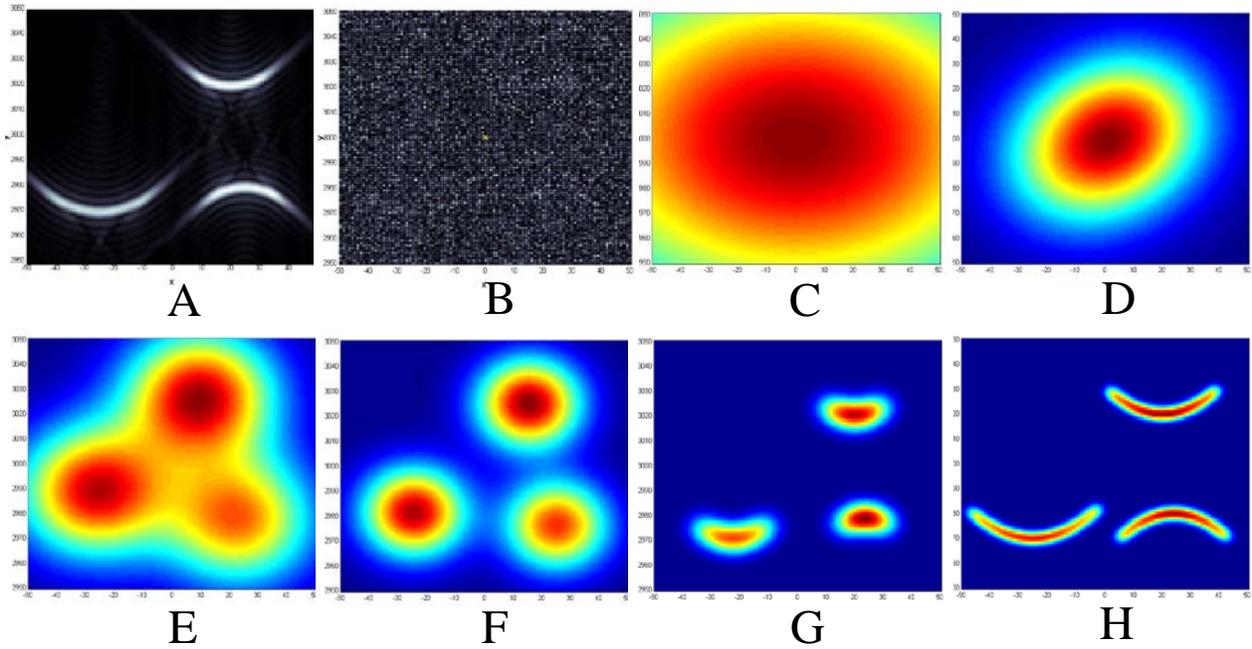

Fig.1. Finding 'smile' and 'frown' patterns in noise, an example of DL operation: (A) true 'smile' and 'frown' patterns are shown without noise; (B) actual image available for recognition (signal is below noise, signal-to-noise ratio is about 1/3 to 1/2 of noise; (C) an initial fuzzy blob-model, the fuzziness corresponds to uncertainty of knowledge; (D) through (H) show improved models at various iteration stages (total of 21 iterations). Between stages (D) and (E) the algorithm tries to fit the data with more than one model and decides, that it needs three blob-models to 'understand' the content of the data. There are several types of models: one uniform model describing noise (it is not shown) and a variable number of blob-models and parabolic models, which number, location, and curvature are estimated from the data. Until about stage (G) the algorithm 'thinks' in terms of simple blob models, at (G) and beyond, the algorithm decides that it needs more complex parabolic models to describe the data. Iterations stopped at (H), when similarity stops increasing. This example is discussed in more details in[10].

### The Knowledge Instinct (KI)

DL is mathematically equivalent to maximizing similarity between mental representations and sensor data about the world. This similarity is a measure of knowledge (of the world). The DL process therefore increases knowledge. Perception, and the very survival of an organism requires constant execution of this process. It is an inborn mechanism and does not depend on our will. In other words, it is an instinct, and it is called the Knowledge Instinct. As discussed below, KI is a foundation of all our higher mental abilities.

The word "emotion" is used to designate many different processes in mind. Here we only consider one mechanism of emotions, described by instinctual-emotional theory of Steven Grossberg and Daniel Levine[11]; emotions are neural signals conveying to decision-making brain areas information about objects and events that could satisfy instinctual needs. For example, if one's instinct for food is not satisfied, one feel hunger. Here we concentrate on KI and related emotions. These emotions related to knowledge are called aesthetic emotions. They are not necessarily related to art; first of all they are related to every process of perception and cognition. Subjectively we perceive them as harmony or disharmony between contents of our mental representations and surrounding world. When the objects around corresponds to our expectations, the emotions of harmony due to correctly perceiving them is below the threshold of consciousness. Similarly, when the stomach properly performs its job, we are not conscious about it. But as soon as stomach fails, we are immediately conscious about related emotions. The same happens with KI, as soon as it is dissatisfied at the most basic levels, when objects do not behave as expected, this is immediately experienced as a disharmony, even as terror. This is a standard content of horror movies: when everyday objects behave unexpectedly, this could be terrifying.

Aesthetic emotions, positive and negative, could be conscious at higher levels of the mind hierarchy. When we have solved a problem that occupied us for a while, KI is satisfied at a higher level and this aesthetic emotion can be consciously experienced. Representations of abstract ideas, at higher levels in the mind, unify knowledge at many lower levels; they are more important, their understanding lead to stronger satisfaction of KI and to stronger aesthetic emotions. Representations at the very top of the mind hierarchy attempt to unify all our knowledge. This unifying understanding is experienced as a purpose and meaning of our existence. Can this really be achieved? Is there a meaning and purpose to human life? We discussed previously that representations of even simple everyday objects are vague. Representations of higher, more abstract objects are built on several layers of vagueness. The higher in the hierarchy the vaguer and less conscious are representations. Their conceptual and emotional contents are not well separated. Representations at the very top of the mind hierarchy are vague. The meaning and purpose of life can never be as clear and conscious as an object in front of our eyes. The life never gives as direct evidence that our lives have a meaning and purpose. However, being sure that one's life is meaningful and purposeful is so important for concentrating will, for survival, for achieving higher goals that any event convincing us in this is experienced as a powerful emotion. According to DL as well as according to Kantian aesthetics, this emotion, satisfying KI at the very top of the mind hierarchy, is the emotion of the beautiful [12,13,14].

## Symbolic culture

Terrence Deacon wrote "Symbol is the most misused word in our culture"[15]. We use this word in trivial cases referring to traffic signs or letters of an alphabet, and in the most profound cases of religious symbols moving entire cultures for millennia. To understand the source of this confusion we need to analyze mechanisms of cognition and language in their interaction.

The interaction of language and cognition in our brain-mind is attained by neural connections between cognitive and language representations. Whereas cognitive representations model objects and events in the world, language representations model sounds, words, phrases, and grammar of the language[16]. Language is learned from surrounding language, where it exists "ready-made"; this is why language can be learned without much life experience by 5 years of age. Learning cognitive representations requires life experience. Experience is not sufficient, because there are always many events of no significance. For example, a particular pattern on the ceiling of a symphony hall is not essential for understanding music. We easily discard these nonessential details, but how do we learn to do it? It is because of the interaction between language and cognition that whenever a child hears a word his KI forces him (or her) to define the contents of the corresponding cognitive representation. Thus cognition is grounded in both experience and language. This is why human cognition is only possible due to human language. The dual model enabling this interaction consists in inborn neural connections between language and cognitive representations. When a baby is born, there is no specific word sounds "chair" and no images of chair in the mind, but neural connections between these future representations are inborn. Gradually, representations are adapted to surrounding language and to experience. Higher in the mind hierarchy, contents of abstract cognitive representations may remain vague and unconscious throughout the life, while using language people can discuss cultural contents crisply and consciously, Fig.2. This is obvious when observing kids, they often talk without full understanding; but the same is true about adults. Sometimes people speak without full understanding. The dual model combined with DL processes, connecting conscious and unconscious, are psychological process-symbols. DL and the dual model enabling these processes are fundamental first principles of human language and cognition.

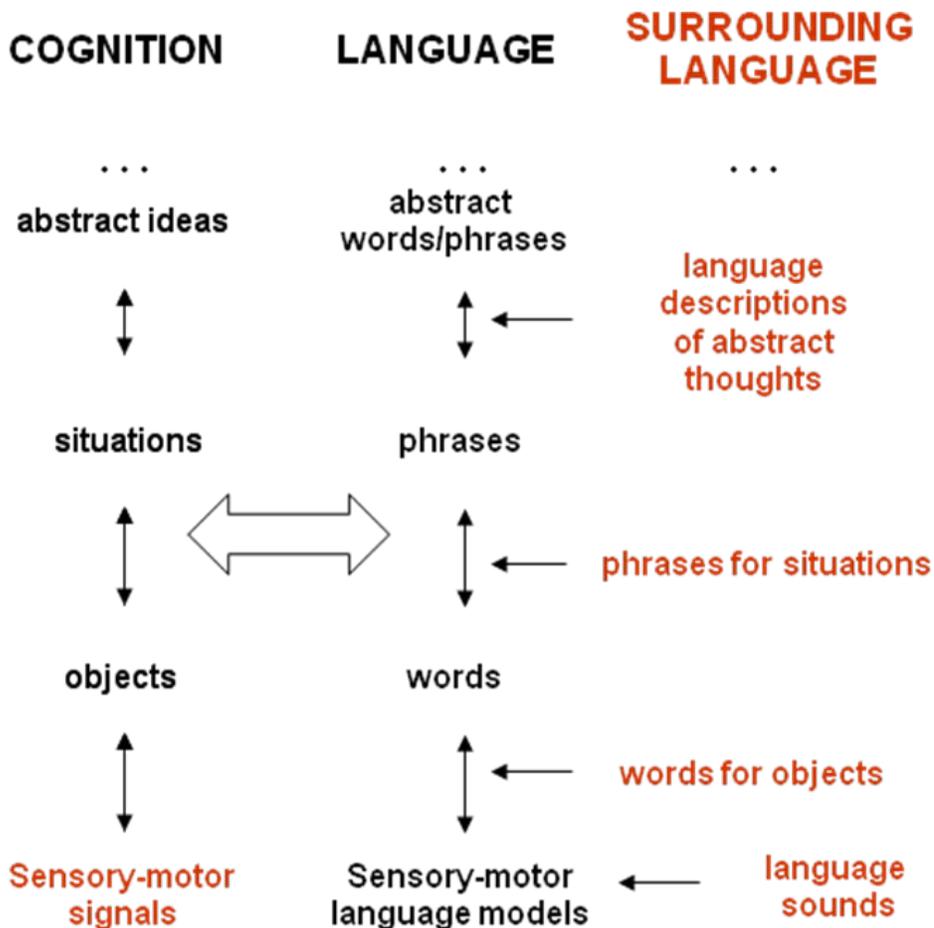

Fig.2. A dual hierarchy of language and cognition. Language is grounded in the surrounding language. Perception is grounded in sensory-motor experience. Higher level cognition is grounded in perception and language.

## Music

According to Darwin[17], music "must be ranked amongst the most mysterious (abilities) with which (man) is endowed." Aristotle listed the power of music among the unsolved problems[1]. The consensus of contemporary scientists is summarized by Masataka[18], "Music is a human cultural universal that serves no obvious adaptive purpose, making its evolution a puzzle for evolutionary biologists".

A possible explanation for this mystery is proposed in[19]. There is no dual model in animals mind; their vocalization and cognition are unified, as well as their conceptual, emotional, and behavioral mental states. When a monkey sees an approaching leopard, it understands this dangerous situation conceptually, it is afraid emotionally, it appropriately behaves by jumping on a tree, and it vocalizes "tiger" in monkey's language so that the rest of the pack would jump on a tree. A monkey experiences a unified conceptual-emotional-behavioral-voicing mental state. A human can say "tiger," without being scared by actual tiger, no animal can voluntary and consciously do it. When human progenitors were separating from animal kingdom due to evolution of language, human brain was rewired, so that emotions and vocalizations separated, also conceptual understanding and behavior separated. This was necessary so that deliberate conversations became possible.

Human progenitors paid heavy price for this freedom, their psyche lost automatic inborn unity and became split. But living with split psyche is impossible; unity of psyche is necessary for concentrating will, for survival. The dual hierarchy of Fig. 2 had to be unified. This required specific motivation or emotion directed at unification. Therefore, while one part of animal voice was losing its emotionality, acquiring semantics, and becoming language, another part of animal voice was increasing its emotionality at the expense of semantics, and was becoming music. Thus cognitive function of music is to maintain unify of psyche, along with acquiring diverse knowledge. Language gave a tremendous evolutionary advantage to those of our progenitors, who could maintain unity of psyche along with differentiated knowledge. It is not easy, because any two pieces of knowledge are contradictory to some extent; even as trivial a contradiction as: "Do you prefer tea or coffee?" Contradictions in knowledge are called cognitive dissonances, resolving them requires variety of emotions. The number of combinations of knowledge pieces is huge, similarly huge is the number of musical emotions. Resolving contradiction in knowledge was not easy when contemporary consciousness was emerging 2,500 years ago in Ancient Greece. Psyche was differentiated and dithyrambs were used to unify psyche. The same is mental function of music today, including pop songs and rap. In style and performance rap is similar to Ancient Greek dithyrambs. In both dithyramb and rap – quite regular thoughts are cried out at the edge of frenzy. As in Ancient Greece 2,500 years ago, so today in a complex multiform culture, people, especially young people, are losing their bearings. Words no longer call forth emotional reactions, their prime emotional meanings are lost. By shouting words along with primitive melody and rhythm, a human being limits his or her conscious world, but restores connection of conscious and unconscious. An internal world comes to wholeness, reunites with a part of the surrounding culture. Music by Bach, Beethoven, and Chopin serves the same cognitive function directed at reconciling profound dissonances in our souls. Music is necessary for continuing evolution of complex cultures.